# Scattering Amplitude Together with Thermodynamic Properties in the Pöschl-Teller Double Ring-Shaped Coulomb Potential

M. Eshghi [1,*], S. Soudi[2], S.M. Ikhdair[3,4]

[1] *Department of Physics, Imam Hossein Comprehensive University, Tehran, Iran*
[2] *Department of Medical Engineering, Faculty of Health and Medical Engineering, Tehran Medical Sciences, Islamic Azad University, Tehran, Iran*
[3] *Department of Physics, Faculty of Science, An-Najah National University, Nablus, West Bank, Palestine*
[4] *Department of Electrical Engineering, Near East University, Nicosia, Northern Cyprus, Mersin 10, Turkey*

## Abstract

We obtain the exact solution to the Dirac equation with the Pöschl-Teller double ring-shaped Coulomb (PTDRSC) potential for any spin-orbit quantum number $\kappa$. The relativistic scattering amplitude for spin $1/2$ particles in the field of this potential has been studied. The wave functions are expressed in terms of the hyper-geometric series of the continuous states on the $k/2\pi$ scale. A formula for the phase shifts has also been found. In the nonrelativistic limits, our solution to the Dirac system converges to that of the Schrödinger one. At the high temperature, the partition function is calculated in order to study the behavior of some thermodynamic properties.

---

[*] **Corresponding Author Email:** *eshgi54@gmail.com*

## 1. Introduction

Solution to relativistic equations plays an important in many aspects of modern physics. In particular, the Dirac equation is the most frequently used wave equation in the description of particle dynamics in relativistic quantum mechanics and in many fields of physics such as nuclear and high-energy physics and chemistry. In recent years, there has been an increase in searching for analytic solution to the Dirac equation; for example, see [1-12]. Also, there has been continuous interest in studying of the scattering states solution within the framework of non-relativistic and relativistic quantum mechanics for central and non-central potentials [13-19]. Because the scattering problems in the presence of an external potential has become interesting topics in relativistic and non-relativistic quantum mechanics and scattering of a relativistic particle by a potential can be treated exactly by finding the continuum solutions of the Dirac equation.

Here, we intend to solve the Dirac equation under the (PTDRSC) potential. This Potential [20] is a kind of physical potential which in spherical coordinates is given by

$$V(r,\theta,\phi) = -\frac{\delta}{r} + \frac{1}{r^2}\left[\frac{B}{\sin^2\theta} + \frac{A(A-1)}{\cos^2\theta}\right] \\ + \frac{1}{r^2\sin^2\theta}\left[\frac{\alpha^2 D(D-1)}{\sin^2\alpha\phi} + \frac{\alpha^2 C(C-1)}{\cos^2\alpha\phi}\right], \qquad (1)$$

where $A, C, D > 1$; $\delta > 0$; $B > 0$; $\alpha = 1, 2, 3, ...$ are real positive parameters. Maghsoodi *et al* have solved the Dirac equation for the above potential with the NU method [21]. In this work, we attempt to investigate the scattering states of the Dirac equation for the above potential and will discuss some of the analytical properties and also calculate for the nonrelativistic particles their thermodynamic properties.

The paper is organized as follows. In section 2, we intend to solve the dirac equation with the Pöschl-Teller double ring-shaped Coulomb (PTDRSC) potential for any spin-orbit quantum number $\kappa$. In section 3, we obtain the continuous energy states and wave functions for the radial and angular parts of the Dirac equation. We consider the limiting case to obtain the nonrelativistic energy solution and then the partition function at high temperature to study the thermodynamic properties of the present model. Finally we end with our discussion and conclusions in section 4.

## 2 The Dirac equation

The Dirac Hamiltonian (in natural units of $\hbar = c = 1$) is [22, 23]

$$H = \vec{\alpha}\cdot\vec{p} + \beta\left(M + S(\vec{r})\right) + V(\vec{r}), \tag{2}$$

where $S(r)$ and $V(r)$ stand for scalar and vector potentials, respectively, $\alpha$ and $\beta$ are Dirac matrices and $M$ denotes the mass. The Dirac equation can be written as

$$\left[\vec{\alpha}\cdot\vec{p} + \beta\left(M + S(\vec{r})\right) + V(\vec{r})\right]\Psi(\vec{r}) = E\,\Psi(\vec{r}), \tag{3}$$

where $E$ denotes the energy and momentum $\vec{p} = -i\nabla$. In Pauli-Dirac representation, because of the appearance of $4 \times 4$ matrices in the Dirac equation, the wave function must be a four-component vector. It is useful to classify the upper two and lower two components of the Dirac wave function as two-component spinors [24]. We write

$$\psi(\vec{r}) = \begin{pmatrix} \varphi^L(r) \\ \chi^S(r) \end{pmatrix} \equiv \begin{pmatrix} \varphi(r) \\ \chi(r) \end{pmatrix}, \tag{4}$$

where $\varphi^L(r)$ and $\chi^S(r)$ are termed the *large* and *small* components of the wave function, we get

$$\vec{\sigma}\cdot\vec{p}\,\chi(\vec{r}) = \left[E - V(\vec{r}) - M - S(\vec{r})\right]\varphi(\vec{r}), \tag{5}$$

$$\vec{\sigma}\cdot\vec{p}\,\varphi(\vec{r}) = \left[E - V(\vec{r}) + M + S(\vec{r})\right]\chi(\vec{r}). \tag{6}$$

When scalar potential is equal to the vector potential, the above equations turn out to become

$$\vec{\sigma}\cdot\vec{p}\,\chi(\vec{r}) = \left[E - M - 2V(\vec{r})\right]\varphi(\vec{r}), \tag{7}$$

$$\chi(\vec{r}) = \frac{\vec{\sigma}\cdot\vec{p}}{E + M}\varphi(\vec{r}). \tag{8}$$

Further, substituting Eq. (8) into Eq. (7), we obtain a Schrodinger-like equation for the upper component

$$\left[p^2 + 2(E + M)V(\vec{r})\right]\varphi(\vec{r}) = \left[E^2 - M^2\right]\varphi(\vec{r}), \tag{9}$$

and substituting potential (1) into Eq. (9), we have

$$\left[-\nabla^2 + 2(E+M)\left(\begin{array}{c}-\dfrac{\delta}{r}+\dfrac{1}{r^2}\left[\dfrac{B}{\sin^2\theta}+\dfrac{A(A-1)}{\cos^2\theta}\right]\\ +\dfrac{1}{r^2\sin^2\theta}\left[\dfrac{\alpha^2 D(D-1)}{\sin^2\alpha\phi}+\dfrac{\alpha^2 C(C-1)}{\cos^2\alpha\phi}\right]\end{array}\right)\right]\varphi(r,\theta,\varphi) \quad (10)$$

$$=\left[E^2-M^2\right]\varphi(r,\theta,\varphi).$$

For a spherical potential, substituting the wave function

$$\varphi_{nlm}(r,\theta,\varphi)=\frac{g_{nlm}(r)}{r}\frac{H_l(\theta)}{(\sin\theta)^{1/2}}\Phi_m(\phi), \quad (11)$$

into Eq.(10) leads to the following set of second-order differential equations

$$\frac{d^2 g(r)}{dr^2}-\left\{\left(\frac{2(M+E)\delta}{r}+\frac{1/4-l^2}{r^2}\right)-M^2+E^2\right\}g(r)=0, \quad (12)$$

$$\frac{d^2 H(\theta)}{d\theta^2}+\left\{\frac{1/4-2(E+M)-m^2}{\sin^2\theta}-\frac{2(E+M)A(A-1)}{\cos^2\theta}+l^2\right\}H(\theta)=0, \quad (13)$$

$$\frac{d^2\Phi(\phi)}{d\phi^2}+\left\{\frac{-2(E+M)\alpha^2 D(D-1)}{\sin^2\alpha\phi}-\frac{2(E+M)\alpha^2 C(C-1)}{\cos^2\alpha\phi}+m^2\right\}\Phi(\phi)=0, \quad (14)$$

where $l=0,1,2,...,$ and $m=0,\pm 1,\pm 2,...$ are separation constants.

## 3.1 Scattering states of the radial Dirac equation

If the energy is positive, then $k=\sqrt{E_{n\kappa}^2-M^2}$ is called the wave number associated to the electron whenever it moves asymptotically for the origin, the center of the force field. Eq. (12) can be written as

$$\frac{d^2 g(r)}{dr^2}+\left[k^2+\frac{2(M+E_{n\kappa})\delta}{r}-\frac{(\ell-1/2)(\ell+1/2)}{r^2}\right]g(r)=0. \quad (15)$$

The boundary conditions for Eq. (15) on the boundaries $g(0)=0$ and $g(\infty)$ are to be finite values. Owing to the asymptotic behavior of the radial wave functions of the continuous states as $r\to 0$, we need to take the wave functions in the form

$$g(r) = A(kr)^{\ell+1/2} \exp(ikr) f(r). \tag{16}$$

On substituting the wave function (16) into Eq. (15), we have

$$\frac{d^2 f}{dr^2} + (2\ell + 1 + 2ikr)\frac{df}{dr} + [2ik(\ell+1/2) + 2(M+E_{n\kappa})\delta] f = 0. \tag{17}$$

Changing to a new variable $z = -2ikr$, Eq. (17) can be simplified

$$\frac{d^2 f}{dz^2} + (2\ell + 1 - z)\frac{df}{dz} + \left[\ell + 1/2 - \frac{i(M+E_{n\kappa})\delta}{k}\right] f = 0, \tag{18}$$

whose analytical solutions as $r \to 0$ are the confluent hypergeometric functions [25, 26]

$$f(r) = F\left(\ell + \frac{1}{2} - \frac{i(M+E_{n\kappa})\delta}{k}, 2\ell + 1, -2ikr\right). \tag{19}$$

Thus, the radial wave function of the scattering states are expressed

$$g_{k\ell}(r) = A_{k\ell}(kr)^{\ell+1/2} \exp(ikr) F\left(\ell + 1/2 - \frac{i(M+E_{n\kappa})\delta}{k}, 2\ell + 1, -2ikr\right). \tag{20}$$

We now study asymptotic form of the above expression for large $r$ to calculate the normalization constant $A_{k\ell}$ of radial wave functions and the phase shifts $\delta'_\ell$. Further, the asymptotic expression of the confluent hypergeometric functions when $|z| \to \infty$ is given by [25, 26]

$$F(\eta, \gamma, z) \to \frac{\Gamma(y)}{\Gamma(\eta)} e^z z^{\eta-\gamma} + \frac{\Gamma(y)}{\Gamma(\gamma-\eta)} e^{\pm i\pi\eta} z^{-\eta}, \tag{21}$$

the upper sign in the second term applies for $-\pi/2 < \arg z < 3\pi/2$ and the lower sign in the second term applies for $-3\pi/2 < \arg z < -\pi/2$, the symbol $\Gamma$ denotes the Gamma function. When $z = -2ikr = |z|e^{-i\pi/2}$, Eq. (21) is then re-expressed as

$$F(\eta, \gamma, z) \to \frac{\Gamma(y)}{\Gamma(\eta)} e^z z^{\eta-\gamma} e^{-i\pi(\eta-\gamma)/2} + \frac{\Gamma(y)}{\Gamma(\gamma-\eta)} e^{-i\pi\eta/2} z^{-\eta}, \tag{22}$$

from which, we have

$$F\left(\ell+\frac{1}{2}-\frac{i(M+E_{n\kappa})\delta}{k},2\ell+1,-2ikr\right)$$

$$\xrightarrow{r\to\infty}\frac{\Gamma(2\ell+1)(2kr)^{-\left(\ell+\frac{1}{2}-\frac{i(M+E_{n\kappa})\delta}{k}\right)}\exp(-2ikr)\exp\left(i\pi\left(\ell+\frac{1}{2}-\frac{i(M+E_{n\kappa})\delta}{k}\right)\Big/2\right)}{\Gamma\left(\ell+\frac{1}{2}-\frac{i(M+E_{n\kappa})\delta}{k}\right)} \tag{23}$$

$$+\frac{\Gamma(2\ell+1)(2kr)^{-\left(\ell+\frac{1}{2}-\frac{i(M+E_{n\kappa})\delta}{k}\right)}}{\Gamma\left(\ell+\frac{1}{2}-\frac{i(M+E_{n\kappa})\delta}{k}\right)}\exp\left(i\pi\left(\ell+\frac{1}{2}-\frac{i(M+E_{n\kappa})\delta}{k}\right)\Big/2\right).$$

If we can write

$$\Gamma\left(\ell+\frac{1}{2}-\frac{i(M+E_{n\kappa})\delta}{k}\right)=\left|\Gamma\left(\ell+\frac{1}{2}-\frac{i(M+E_{n\kappa})\delta}{k}\right)\right|\exp(i\delta_\ell), \tag{24}$$

then

$$\Gamma\left(\ell+\frac{1}{2}+\frac{i(M+E_{n\kappa})\delta}{k}\right)=\left|\Gamma\left(\ell+\frac{1}{2}-\frac{i(M+E_{n\kappa})\delta}{k}\right)\right|\exp(-i\delta_\ell), \tag{25}$$

where $\delta_\ell$ is a real number. Eq. (23) then becomes

$$F\left(\ell+\frac{1}{2}-\frac{i(M+E_{n\kappa})\delta}{k},2\ell+1,-2ikr\right)$$

$$\approx\frac{\Gamma(2\ell+1)(2ikr)^{-\left(\ell+\frac{1}{2}+\frac{i(M+E_{n\kappa})\delta}{k}\right)}}{\left|\Gamma\left(\ell+\frac{1}{2}-\frac{i(M+E_{n\kappa})\delta}{k}\right)\right|}+\frac{\Gamma(2\ell+1)\exp(-2ikr)(2kr)^{-\left(\ell+\frac{1}{2}+\frac{i(M+E_{n\kappa})\delta}{k}\right)}}{\left|\Gamma\left(\ell+\frac{1}{2}+\frac{i(M+E_{n\kappa})\delta}{k}\right)\right|}, \tag{26.a}$$

then

$$F\left(\ell + \frac{1}{2} - \frac{i(M + E_{n\kappa})\delta}{k}, 2\ell + 1, -2ikr\right)$$

$$= \frac{\Gamma(2\ell+1)\exp(-ikr)\exp\left(-\pi(M+E_{n\kappa})\delta/2k\right)}{\left|\Gamma\left(\ell + \frac{1}{2} - \frac{i(M+E_{n\kappa})\delta}{k}\right)\right|(2kr)^{\ell+1/2}}$$

$$\times \left[ (-i)^{-\ell-1/2} \exp\left(-i\left(kr + \delta_\ell - \frac{\ell\pi}{2} + \frac{(M+E_{n\kappa})\delta \ln(2kr)}{k}\right)\right) \right.$$

$$\left. - i^{-\ell-1/2} \exp\left(i\left(kr + \delta_\ell - \frac{\ell\pi}{2} + \frac{(M+E_{n\kappa})\delta \ln(2kr)}{k}\right)\right) \right].$$

(26.b)

When $r \to \infty$, we have

$$F\left(\ell + \frac{1}{2} - \frac{i(M + E_{n\kappa})\delta}{k}, 2\ell + 1, -2ikr\right)$$

$$\xrightarrow{r \to \infty} \frac{\Gamma(2\ell+1)\exp(-ikr)\exp\left(-\pi(M+E_{n\kappa})\delta/2k\right)}{\left|\Gamma\left(\ell + \frac{1}{2} - \frac{i(M+E_{n\kappa})\delta}{k}\right)\right|(2kr)^{\ell+1/2}}$$

$$\times \left[ i\exp\left(-i\left(kr + \delta_\ell - \frac{\ell\pi}{2} + \frac{(M+E_{n\kappa})\delta \ln(2kr)}{k}\right)\right) \right.$$

$$\left. - i\exp\left(i\left(kr + \delta_\ell - \frac{\ell\pi}{2} + \frac{(M+E_{n\kappa})\delta \ln(2kr)}{k}\right)\right) \right].$$

(26.c)

Substituting Eq. (26) into Eq. (20) leads to

$$g_{k\ell}(r) \xrightarrow{r \to \infty} \frac{2A_{k\ell}\Gamma(2\ell+1)\exp\left(-\pi(M+E_{n\kappa})\delta/2k\right)}{\left|\Gamma\left(\ell + \frac{1}{2} - \frac{i(M+E_{n\kappa})\delta}{k}\right)\right|}$$

$$\times \sin\left(kr + \delta_\ell - \frac{\ell\pi}{2} + \frac{\pi}{4} + \frac{(M+E_{n\kappa})\delta \ln(2kr)}{k}\right).$$

(27)

In terms of the following asymptotic behavior

$$g_{k\ell}(r) \xrightarrow{r \to \infty} 2\sin\left(kr + \delta_\ell - \frac{\ell\pi}{2} + \frac{\pi}{4} + \frac{(M+E_{n\kappa})\delta \ln(2kr)}{k}\right),$$

(28)

It is given in Ref. [27, 28] that the radial wave functions of the continuous states for the Coulomb potential are normalized on the "$k/2\pi$ scale". Because the new model potential is a short distance potential in the Poschl-Teller Double Ring-Shaped Coulomb Potential, so it has no influence on asymptotic expression of the wave function for large $r$.

It is useful to note that, considering the asymptotic behavior of the wave function, the scattering amplitudes are also valid if we take into account the relativity [29, 30]. In other words, the asymptotic expression of the Pöschl-Teller Double Ring-Shaped Coulomb potential is identical to that of the Coulomb potential when $r \to \infty$, i.e.

$$g_{k\ell}(r) \xrightarrow{r \to \infty} 2\sin\left(kr + \delta'_\ell - \frac{\ell\pi}{2} + \frac{\pi}{4} + \frac{(M+E_{n\kappa})\delta \ln(2kr)}{k}\right). \tag{29}$$

The wave functions of the continuous states for the Pöschl-Teller double ring-shaped Coulomb potential are normalized on the "$k/2\pi$ scale", too. Here $\delta'_\ell$ represents the phase shifts. If we compare Eq. (29) with Eq. (23), we may obtain the normalization constant of the continuous states as

$$A_{k\ell} = \frac{2^{\ell+\frac{1}{2}}\left|\Gamma\left(\ell+\frac{1}{2} - \frac{i(M+E_{n\kappa})\delta}{k}\right)\right|\exp\left(\pi(M+E_{n\kappa})\delta/2k\right)}{\Gamma(2\ell+1)}, \tag{30}$$

and the phase shifts $\delta'_\ell$ for a short ranged interaction. It is the additional $\ell'$-independent phase shift, $-(M+E_{n\kappa})\delta \ln(2kr)/k$, that distinguishes the Coulomb-like solution from that for a short ranged potential and can be calculated explicitly:

$$\delta'_\ell = \delta_\ell + \pi(\ell' - \ell + 1/2)/2 = \arg\Gamma(\ell+1/2 - i(M+E_{n\kappa})\delta/k) + \pi(\ell' - \ell + 1/2)/2. \tag{31}$$

Substituting Eq. (30) into (20), the normalized wave functions of the continuous states on the "$k/2\pi$ scale" are

$$g_{k\ell}(r) = \frac{2^{\ell+1/2}\left|\Gamma\left(\ell+\frac{1}{2} - \frac{i(M+E_{n\kappa})\delta}{k}\right)\right|\exp\left(\pi(M+E_{n\kappa})\delta/2k\right)}{\Gamma(2\ell+1)}$$
$$\times \exp(ikr)(kr)^{\ell+1/2} F\left(\ell + \frac{1}{2} - \frac{i(M+E_{n\kappa})\delta}{k}, 2\ell+1, -2ikr\right), \tag{32}$$

where $k = \sqrt{E_{n\kappa}^2 - M^2}$.

Before concluding this section, let us study the properties of the scattering amplitude. As we know, once the phase shifts are obtained, we can study the scattering amplitude and the differential cross section. For the sake of simplicity, following [31] we can obtain the scattering amplitude as

$$f(\theta) = -\frac{i}{\sqrt{2\pi k}} \sum_{\ell} [\exp(2i\delta_\ell) - 1] e^{im\theta}, \tag{33}$$

due to $\sum_\ell e^{im\theta} = 2\pi\delta(\theta)$, we have for $\theta \neq 0$

$$f(\theta) = -\frac{i}{\sqrt{2\pi k}} \sum_{\ell} [\exp(2i\delta_\ell)] e^{im\theta}, \tag{34}$$

from which we may calculate the cross section

$$\sigma(\theta) = |f(\theta)|^2 = \left| -\frac{i}{\sqrt{2\pi k}} \sum_{\ell} [\exp(2i\delta_\ell)] e^{im\theta} \right|^2. \tag{35}$$

It should be noted that it is very difficult to obtain an analytical expression for Eq. (28). Nevertheless, Dong and Lozada-Cassou [32] have obtained cross section in the special case as

$$\sigma(\theta) = |f(\theta)|^2 = \left| -i \frac{\Gamma(1/2 - i\alpha) e^{i\alpha \ln(\sin^2(\theta/2))}}{\sqrt{2k}\Gamma(i\alpha)\sin(\theta/2)} \right|^2 \tag{36}$$

$$\underbrace{\Gamma(iy)\Gamma(-iy) = |\Gamma(iy)|^2 = \frac{\pi}{y\sinh(\pi y)} \quad \text{and} \quad \Gamma\left(\frac{1}{2}+iy\right)\Gamma\left(\frac{1}{2}-iy\right) = \left|\Gamma\left(\frac{1}{2}+iy\right)\right|^2 = \frac{\pi}{y\cosh(\pi y)}} \to \sigma(\theta) = \frac{\alpha \tanh(\pi\alpha)}{2k \sin^2(\theta/2)}.$$

For more information about scattering, see Appendix A. In here, we shall discuss the analytical properties of the scattering amplitude in the entire complex $k$ plane by regarding the scattering amplitude as the function of the phase shifts. To this end, from formula (35), we need discuss analytical properties of $\Gamma(\ell + 1/2 - i(M + E_{n\kappa})\delta/k)$. The Gamma function $\Gamma(z)$ has simple poles at $z = 0, -1, -2, \ldots$. To see this we can use to write

$$\Gamma(z) = \frac{\Gamma(z+1)}{z} = \frac{\Gamma(z+2)}{z(z+1)} = \frac{\Gamma(z+3)}{z(z+1)(z+2)} = \ldots. \tag{37}$$

Clearly Gamma function, $\Gamma(z)$, has a pole at $z = 0$ with residue, at $z = -1$ with residue, at $z = -2$ with residue, *etc*. Also $\Gamma(z)$ is never zero in the complex plane. Namely, the first order poles of $\Gamma(\ell + 1/2 - i(M + E_{n\kappa})\delta/k)$ is situated at

$$\ell + 1/2 - i(M + E_{n\kappa})\delta/k = 0, -1, -2, \ldots = -n_r, \qquad n_r = 0, 1, 2, \ldots . \tag{38}$$

At these poles, the corresponding energy levels are given by

$$(2n_r + 1)\sqrt{M^2 - E^2} - 2(E + M)\delta + 2\sqrt{\ell^2(M^2 - E^2)} = 0 \tag{39}$$

where Eq. (39) is the same with Eq. (21) of Ref. [21]. The energy equation (39) gives the bound states for the PTDRSC potential as reported before in [21].

Let us find the non-relativistic solution, by using the following transformations $E_{nl} \approx E - M$, $E + M \approx 2\mu$. The relativistic energy equation (39) reduces to the well-known binding energy of the Coulomb potential, that is, $E_n = -\mu(2n + 2\ell + 1)^2 / 2\delta^2$.

Having calculated the energy, we can investigate the thermodynamics properties [33]. Thus, we can immediately obtain the thermodynamics quantities of the system in a systematic manner. In order to obtain all thermodynamic quantities of the non-relativistic particles system, we should concentrate on the calculation of the canonical partition function Z. The partition function Z at finite temperature T, is obtained through the Boltzmann factor as $Z = \sum_{n=0}^{-\ell-1/2} e^{-\beta E_{n\ell}}$ where $\beta = 1/k_B T$ and $k_B$ is the Boltzmann constant. The partition function Z be as

$$Z = \sum_{n=0}^{-\ell-1/2} \exp\left(\frac{n + \ell + 1/2}{\delta/\sqrt{2\beta\mu}}\right)^2 . \tag{40}$$

In the classical limit, at high temperature $T$ for large $\xi$ and small $\beta$, the sum can be replaced by the following integral

$$Z = \gamma \int_0^\xi e^{y^2} dy = \frac{\tau\sqrt{\pi}\, Erfi\left(\frac{\xi}{\tau}\sqrt{\beta}\right)}{2\sqrt{\beta}}, \tag{41}$$
10

where $y = (n - \xi/\gamma)$, $\xi = -\ell - 1/2$, $\gamma = \delta/\sqrt{2\beta\mu}$ and $\tau = (\delta/\sqrt{2\mu\beta})$. The imaginary error function is an entire function defined by $Erfi(x) = iErf(ix)$, where $erf$ denotes the error function is a special function of sigmoid shape which occurs in probability, statistics and partial differential equations. In mathematics, the error function can be denoted as [33, 34] $Erf(x) = (2/\sqrt{\pi})\int_0^x e^{-t^2} dt$. The imaginary error function is implemented in mathematica as $Erfi[x]$. Other thermodynamic properties of the system can be easily obtained from partition function. In fact, any other parameter that might contribute to the energy should also appear in the argument of $Z$ [35]. The vibrational mean energy $U$ is

$$U = -\frac{\partial}{\partial \beta} \ln Z = \frac{1}{2\beta}\left[1 - \frac{\frac{\xi}{\tau}\sqrt{\beta}}{DawsonF(\frac{\xi}{\tau}\sqrt{\beta})}\right]$$

$$= \frac{2\frac{\xi}{\tau}\sqrt{\beta}}{\sqrt{\pi}Erfi\left(\frac{\xi}{\tau}\sqrt{\beta}\right)}\left[\frac{e^{\left(\frac{\xi}{\tau}\sqrt{\beta}\right)^2}}{2\beta} - \frac{\sqrt{\pi}Erfi\left(\frac{\xi}{\tau}\sqrt{\beta}\right)}{4\beta\frac{\xi}{\tau}\sqrt{\beta}}\right],$$
(42)

which implies that $U = -\xi^2/3\tau^2$ when $\beta \ll 1$. In Eq. (42), the Dowson function [33, 34] or Dowson integral can be denoted $F(x) = e^{-x^2}\int_0^x e^{-y^2} dy = (\sqrt{\pi}/2)e^{-x^2} erfi(x)$. Thus the Dowson's integral is implemented in mathematica as $DowsonF[x]$. The vibrational specific heat $C$ is

$$C = -\frac{\partial U}{\partial T} = -k\beta^2 \frac{\partial U}{\partial \beta}$$

$$= \frac{k}{2}\left[1 - \frac{\frac{\xi}{\tau}\sqrt{\beta}\left[\frac{2\xi\sqrt{\beta}}{\tau}e^{\left(\frac{\xi}{\tau}\sqrt{\beta}\right)^2} + \sqrt{\pi}\left(1 - 2\left(\frac{\xi}{\tau}\sqrt{\beta}\right)^2\right)Erfi\left(\frac{\xi}{\tau}\sqrt{\beta}\right)\right]}{4e^{\left(\frac{\xi}{\tau}\sqrt{\beta}\right)^2} DawsonF\left(\frac{\xi}{\tau}\sqrt{\beta}\right)^2}\right],$$
(43)

which yields $C = 0$ when $\beta \ll 1$. The vibrational mean free energy $F$ is

$$F = -kT\ln Z = -\frac{1}{\beta}\ln\left(\frac{\tau\sqrt{\pi}Erfi\left(\frac{\xi}{\tau}\sqrt{\beta}\right)}{2\sqrt{\beta}}\right).$$
(44)

We write a description for reminder about entropy then calculate the entropy from the partition function. Entropy is a fundamental measure of information content which has been applied in a wide variety of fields. Entropy plays an important role in thermodynamics. It is central to the Second Law of Thermodynamics. It helps measure the amount of order and disorder and/or chaos. Entropy can be defined and measured in many other fields than the thermodynamics. For instance, in classical physics, entropy is defined as the quantity of energy incapable of physical movements. In here, the vibrational entropy $S$ that is as

$$S = k \ln Z + kT \frac{\partial}{\partial T} \ln Z_{vib} = k \ln Z - k\beta \frac{\partial}{\partial \beta} \ln Z$$

$$= \frac{k}{2} \left[ 1 - \frac{\frac{\xi}{\tau}\sqrt{\beta}}{DawsonF(\xi)} + 2\log\left(\frac{\tau.Erfi\left(\frac{\xi}{\tau}\sqrt{\beta}\right)}{\sqrt{\beta}}\right) + \log\left(\frac{\pi}{4}\right) \right]. \tag{45}$$

## 3.2. Polar and azimuthal solutions

For this case, by choosing

$$-\frac{1/4 - 2(E + M) - m^2}{\varsigma^2} = \chi(\chi - 1),$$

$$-\frac{2(E + M)A(A - 1)}{\varsigma^2} = \lambda(\lambda - 1), \quad \ell^2/2 = E + \frac{1}{2}\varsigma^2, \tag{46}$$

we can write Eq. (13) as

$$-\frac{1}{2}\frac{d^2 H(q)}{dq^2} + \frac{\varsigma^2}{2}\left\{\frac{\chi(\chi-1)}{\sin^2(\varsigma q)} + \frac{\lambda(\lambda-1)}{\cos^2(\varsigma q)} - \left(E + \frac{1}{2}\varsigma^2\right)\right\} H(q) = 0. \tag{47}$$

Equation (47) is obviously a standard one-dimensional form of the Schrodinger equation with a generalized Poschl-Teller effective potential which admits an exact solution of the form

$$E_{n_r} = \frac{\varsigma^2}{2}(\chi + \lambda + 2n_r)^2,$$

$$H(q) = C \sin^\chi(\varsigma q) \cos^\lambda(\varsigma q) \times {}_2F_1\left(-n_r, \chi + \lambda + n_r, \chi + \frac{1}{2}; \sin^2(\varsigma q)\right), \tag{48}$$

with $\chi, \lambda > 1$, $H(0) = 0$ and $H(\pi/2\varsigma) = 0$, as reported by Salem and Montemayor, Eq. (4.7) in [36]. However, for $\chi = 0,1$ the effective potential (47) collapses into

$$V_{eff}(q(r)) = \frac{\varsigma^2}{2} \frac{\lambda(\lambda-1)}{\cos^2(\varsigma q)}, \qquad (49)$$

which admits an exact solution

$$E_{n_r} = 2\varsigma^2 \left(\frac{\lambda}{2} + n_r\right)^2 - \frac{\varsigma^2}{2},$$

$$H(q) = A \cos^\lambda(\varsigma q) \times {}_2F_1\left(-n_r, \lambda + n_r, \frac{1}{2}; \sin^2(\varsigma q)\right). \qquad (50)$$

We can also obtain exact solution of the Eq. (13) by using the NU method [37], the analytical exact solution of Eq. (13) has been given in Ref. [21] by NU method. To obtain a solution of Eq. (14) and to avoid repetition in our solution, If we substitute $-2(E+M)\alpha^2 D(D-1) = \chi(\chi-1)$, $-2(E+M)\alpha^2 C(C-1) = \lambda(\lambda-1)$ and $m^2 = E + 1/2\varsigma^2$, then Eq. (14) turns to Eq. (47). Using the similar procedure like the ones in above subsection 3.2., eigenvalues and eigenfunctions of the Eq. (14) can be easily obtained. The analytical exact solution of Eq. (14) has also been given in Ref. [21] by NU method.

## 4. Discussion and Conclusions

In this work, we have investigated the solution of the Dirac equation for particles with spin 1/2 in the PTDRSC potential. The continuous energy states of the Dirac equation with this potential have been presented for any spin-orbit quantum number $\kappa$. The wave functions have been expressed in terms of the hyper-geometric series of the continuous states on the $k/2\pi$ scale. Also, formula of the phase shifts was calculated. We recovered the nonrelativistic solutions in the limiting case. We also presented some of the analytical scattering amplitude and thermodynamic properties for nonrelativistic system.

## Appendix A: Review to Scattering Cross Section

Once the differential scattering cross-section is known, the total scattering cross-section $\sigma_{tot}$ and the first transport scattering cross section $\sigma_{tr}$ may be calculated by $\sigma_{tot} = \int (d\sigma/d\Omega) d\Omega$ and $\sigma_{tr} = \int (1 - \cos\theta)(d\sigma/d\Omega) d\Omega$. We can also obtain the ratio $\Xi$ between the transport and the total scattering cross section as $\Xi = \sigma_{tr}/\sigma_{tot}$. A quantity which is very important for describing the scattering processes in the interaction of electrons with the matter, is the already mentioned transport cross section $\sigma_{tr}$. It is related to the mean number of wide-angle collisions $\nu$ defined as $\nu = NR\sigma_{tr}$, where $N$ is the number of molecules per unit of volume in the target and $R$ the maximum range of penetration. It is now easy to calculate the probability of scattering into angular rang from 0 to $\theta$, that is given by $P(\theta) = (2\pi/\sigma_{tot}) \int_0^\theta (d\sigma/d\Omega) \sin\vartheta . d\vartheta$. Other useful quantities that can be given by simple closed formulas are the probability of forward scattering as $P_F = (2\pi/\sigma_{tot}) \int_0^{\pi/2} (d\sigma/d\Omega) \sin\vartheta . d\vartheta$, and the probability of backscattering as $P_B = (2\pi/\sigma_{tot}) \int_{\pi/2}^{\pi} (d\sigma/d\Omega) \sin\vartheta . d\vartheta$. The knowledge of the forward and of the backscattering probabilities allows us to calculate the backscattering coefficient $r(E_0)$ [38-40]. For low atomic number elements and for some oxides, the differential scattering cross section can be approximated by the function as $d\sigma/d\Omega = \Phi/(1 - \cos\theta + \Gamma)^2$, where the quantities $\Phi$ and $\Gamma$ have to be determined in order to obtain the best fit of the total and transport scattering cross section [38].